\documentclass[12pt]{iopart}
\usepackage{iopams}
\expandafter\let\csname equation*\endcsname\relax
\expandafter\let\csname endequation*\endcsname\relax
\usepackage[fleqn]{amsmath}
\usepackage{amssymb}
\usepackage{amsbsy}
\usepackage{color}
\usepackage{graphicx}
\usepackage{subfigure}
\usepackage{epstopdf}
\usepackage{cite}

\usepackage[breaklinks=true,colorlinks=true,linkcolor=blue,urlcolor=blue,citecolor=blue]{hyperref}

\usepackage{appendix}
\usepackage{multirow}

\begin{document}

\title{Optimizing orbits for TianQin}

\author{Bo-Bing Ye$^{1,2}$, Xuefeng Zhang$^{1,2}$, Ming-Yue Zhou$^{3}$, Yan Wang$^{3}$, Hui-Min Yuan$^{1,2}$, Defeng Gu$^{1,2}$, Yanwei Ding$^{1,2}$, Jinxiu Zhang$^{1,2}$, Jianwei Mei$^{1,2}$, Jun Luo$^{1,2}$}

\address{$^1$ TianQin Research Center for Gravitational Physics, Sun Yat-sen University, Zhuhai 519082, China}

\address{$^2$ School of Physics and Astronomy, Sun Yat-sen University, Zhuhai 519082, China}

\address{$^3$ MOE Key Laboratory of Fundamental Physical Quantities Measurements, \\
Hubei Key Laboratory of Gravitation and Quantum Physics, \\
School of Physics, Huazhong University of Science and Technology, Wuhan 430074, China}

\ead{\mailto{zhangxf38@sysu.edu.cn}}



\begin{abstract}
TianQin is a geocentric space-based gravitational-wave observatory mission consisting of three drag-free controlled satellites in an equilateral triangle with an orbital radius of $ 10^{5}$ km. The constellation faces the white-dwarf binary RX J0806.3$\texttt{+} $1527 located slightly below the ecliptic plane, and is subject to gravitational perturbations that can distort the formation. In this study, we present combined methods to optimize the TianQin orbits so that a set of 5-year stability requirements can be met. Moreover, we discuss slow long-term drift of the detector pointing due to orbital precession, and put forward stable orbits with six other pointings along the lunar orbital plane. Some implications of the findings are pointed out. 
\end{abstract}

\vspace{2pc} \noindent{\it Keywords}: space gravitational-wave detection, TianQin, geocentric orbit, orbit design, orbit optimization

\section{Introduction}\label{sec:intro}

\begin{figure} \label{fig_tqsc}
\begin{center}
\includegraphics[width=13cm,height=6cm]{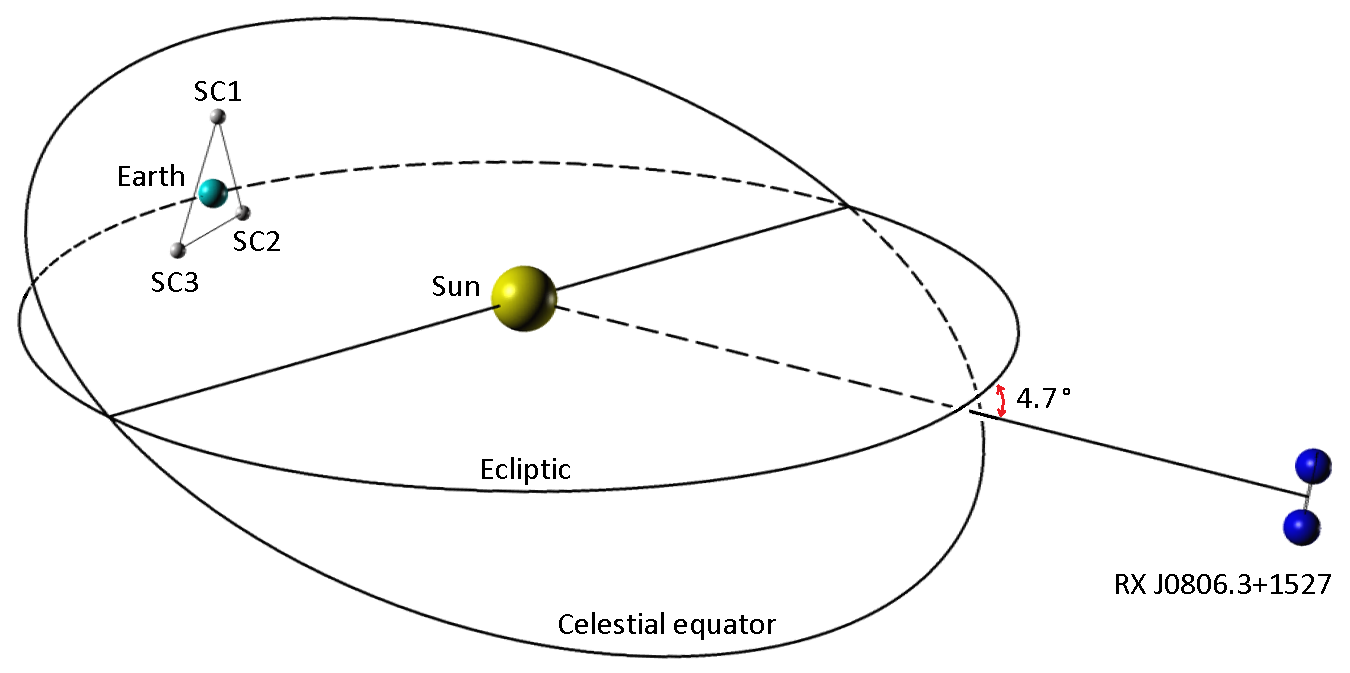}
\caption{An illustration of the TianQin constellation consisting of three spacecraft SC1, SC2, and SC3 (figure reproduced from \cite{Luo2016}). The direction to J0806 is shown. }
\end{center}
\end{figure}

TianQin is a proposed space-borne science mission to detect gravitational waves (GW) in the mHz frequency band \cite{Luo2016}. The mission concept relies on a constellation of three identical drag-free controlled spacecraft in high Earth orbits at an altitude about $ 10^{5} $ km (figure \ref{fig_tqsc}). The constellation forms a nearly equilateral triangle, and the nominal orbital plane stands almost perpendicular to the ecliptic, facing the white-dwarf binary RX J0806.3$ \texttt{+} $1527 (also known as HM Cancri, hereafter J0806) as a reference source \cite{Stroeer2006}. The designed all-sky detection ability of TianQin engenders rich science prospects for GW physics and astronomy \cite{Hu2017}. 

In space, perturbing gravitational forces induce changes in the detector's arm lengths and subtended angles. The science performance must take into account the resulting Doppler shift in laser signals and the beam pointing variations. Therefore, careful orbit analysis and optimal design are necessary to alleviate demands on on-board instruments.

At the geocentric distance of $ 10^{5}$ km, the gravitational perturbations to the TianQin spacecraft are primarily caused by, in descending order of magnitudes, the Moon, the Sun, and the Earth's $ J_{2} $ oblateness. Their relative magnitudes are, respectively, of $ 4\times 10^{-4} $, $ 2\times 10^{-4} $, and $ 6\times 10^{-6} $, compared with the central force from the Earth \cite{Montenbruck2001}. Hence, the Moon constitutes the largest factor affecting the formation stability.

In regard to orbit stability and optimization for other space GW detection missions, the well-known heliocentric LISA design \cite{Vincent1987,Folkner1997,Sweetser2005} has been extensively studied with analytic and numerical methods \cite{Dhurandhar2005,Nayak2006,Marchi2012,Yi2008,Hughes2002,Povoleri2006,Li2008,Xia2010,Halloin2017,Wang2013a}. Particularly, the cost-function method based on carefully chosen performance measures has proved effective in numerical optimization \cite{Hughes2002,Povoleri2006,Li2008,Xia2010,Halloin2017}. For the ASTROD-GW mission \cite{Ni2013}, optimized orbits near the Sun-Earth Lagrange points (L3, L4, and L5) have been acquired through tuning the average periods and eccentricities of the orbits \cite{Men2010,Wang2013b,Wang2015,Ni2016}. Our optimization scheme benefits from these studies. 

An example of simulated TianQin orbits has been given in \cite{Luo2016} (also \cite{Hu2015}), however, without providing much details and further discussion. In this study, we intend to fill in the blanks through an independent verification, and extend the current understanding in TianQin's geocentric orbit design. The paper is organized as follows. In section 2, we introduce the nominal circular orbits of TianQin and preliminary stability requirements on constellation geometry (estimates on eccentricities given in \ref{app:ecc}). In section 3, we describe the simulation tool and initial setup for orbit integration. In section 4, the methods and three steps of optimization are presented (detailed derivation deferred to \ref{app:eq_rv} and \ref{app:eq_io}). The optimized TianQin orbits are shown in section 5, and in section 6, we analyze observed secular pointing drift from the reference source. In section 7, we present the optimization results for other orbital pointings. The paper concludes in section 8.


\section{TianQin mission requirements}

The nominal orbits of the TianQin constellation can be readily given in terms of Kepler's circular orbits. For a design baseline, the three spacecraft have the same orbit radius and form an equilateral formation revolving around the Earth. They fly in the same orbital plane oriented constantly towards the designated reference source J0806. Using the J2000-based Earth-centered ecliptic coordinate system (EarthMJ2000Ec), we prescribe the nominal orbits as follows\footnote{In the geocentric equatorial coordinate system, one has $i=74.541\,611^\circ$ and $\Omega=211.596\,667^\circ$.} \cite{Luo2016}:
\begin{equation} \label{001} 
\eqalign{
a = 10^5\ \mathrm{km},  \cr
e = 0,  \cr
i = 94.704\,035^\circ,  \cr
\Omega = 210.443\,557^\circ,  \cr
\omega = 0^\circ,  \cr
\nu_k = \nu_1 + (k-1)\cdot 120^\circ, \qquad k=1,2,3. 
}
\end{equation}
The orbital elements listed above include the semi-major axis $a$, eccentricity $e$, inclination $i$, longitude of ascending node $\Omega$, argument of periapsis $\omega$, and true anomaly $\nu_k$ ($\nu_1$ for the spacecraft SC1, etc.). The geocentric formation travels in prograde motion with a period of $\sim 3.6$ days. Given the inclination $i$, the nominal orbital plane stands almost upright to the ecliptic plane, which helps reduce direct sunlight into the optical assemblies. Moreover, the normal of the triangle is aiming at J0806. The fixed detector pointing of the nominal constellation differs greatly from the LISA orbit design which features a yearly pointing variation.

Due to various perturbing forces in space, real-world orbits deviate from the simple Keplerian approximation. The arm lengths, relative line-of-sight velocities (range rates), and breathing angles (between arms) of the spacecraft formation, as well as the orbital planes, undergo continuous changes. To accommodate instrumentation and science operations such as Doppler measurement, telescope steering, and drag-free control on board the spacecraft \cite{Folkner1997}, the orbits should be designed to meet certain stability requirements. For this study alone, we have assumed the following permissible variations of constellation geometry in table \ref{tab_req}~\cite{Luo2016}. Note that no requirement is imposed on pointing stability at J0806 since we expect it to be non-critical to the mission (in fact, larger pointing variations benefit sky localization of GW sources). 
\begin{table}[!ht]
\caption{\label{tab_req} 5-year formation stability requirements assumed in this study.}
\begin{indented}
\item[]\begin{tabular}{@{}lll}
\br
Parameter & Permitted range \\ 
\mr
Arm length $L_{ij}$ & $\pm 1 \%\times (\sqrt{3}\times 10^5)\ \mathrm{km} $  \\
Relative velocity $v_{ij}$ & $\pm 10\ \mathrm{m/s}$ \textrm{for 5 years} \\
 & $\pm 5\ \mathrm{m/s}$ \textrm{for the first 2 years} \\
Breathing angle $\alpha _{i}$ & $\pm 0.2^\circ$ \textrm{for 5 years} \\
 & $\pm 0.1^\circ$ \textrm{for the first 2 years} \\
\br
\end{tabular}
\end{indented}
\end{table}


\section{Orbit propagation}

The orbit propagation is implemented by the NASA General Mission Analysis Tool (GMAT)~\cite{GMAT}, which is an open-source, flight qualified software extensively used for space mission design \cite{Hughes14}. The force models we have adopted include a $10\times 10$ spherical-harmonic model of the Earth's gravity field (JGM-3 \cite{Tapley1996}), the point-mass gravity field from the Moon, Sun and solar system planets (the ephemeris DE421 \cite{Folkner2008}), and the first-order relativistic correction\footnote{Other small effects, such as higher-order ($>10$) Earth gravity, the Earth tides, the non-spherical gravity of the Moon and the Sun, have been tested or estimated to be negligible to the optimization results (Table \ref{tab_all}), hence not included for the sake of computational efficiency. }. As the spacecraft are drag-free controlled, we only consider, for a design baseline, purely gravitational trajectories and assume no orbit correction maneuvers performed during the formation flight. A ninth-order Runge-Kutta integrator with eighth-order error control (RungeKutta89) is used. GMAT also provides an optimization solver (fmincon) through an interface with MATLAB.

Under planetary perturbation, orbit propagation from the initial elements provided by the nominal orbits, unless under fortuitous circumstances, fails to satisfy the stability requirements. Generally in these cases, a long-term linear drift in arm lengths and breathing angles can be observed, indicating one spacecraft chasing another within the constellation (see, e.g., \cite{Men2010,Wan2017}). Therefore, one needs to adjust the initial orbital elements (positions and velocities), as free variables, to stabilize the ensuing relative orbital motion. The nominal orbits provide a suitable initial guess for such an optimization procedure. Hence we have assumed the initial elements and epoch (22 May, 2034 12:00:00 UTC, for testing optimization only) in table \ref{tab_nom}. 

\begin{table}[!ht]
\caption{\label{tab_nom} The initial elements from the nominal orbits of the TianQin constellation in the J2000-based Earth-centered ecliptic coordinate system (EarthMJ2000Ec) at the epoch 22 May, 2034 12:00:00 UTC.}
\begin{indented}
\item[]\begin{tabular}{@{}lllllll}
\br
  & $  a $\,(km) &  $  e $ & $ i $\,($ ^{\circ} $) & $ \Omega $\,($ ^{\circ} $) & $  \omega   $\,($ ^{\circ} $)  & $ \nu $\,($ ^{\circ} $) \\ 
\mr
SC1, 2, 3 & $ 10^{5} $  & 0 &  94.704\,035 & 210.443\,557 & 0  & 60, 180, 300 \\
\br
\end{tabular}
\end{indented}
\end{table}


\section{Optimization method}

The optimization starts with orbit propagation from a set of initial elements $\sigma_0 \equiv (a_0, e_0, i_0, \Omega_0, \omega_0, \nu_0) $ (or equivalently, initial positions and velocities) at an epoch $t_0$, which is usually derived from nominal orbits. Based on the resulting orbital behavior, one can make adjustment to $\sigma_0$ and test new orbital elements $\sigma _0'$ from the same $t_0$, and then repeat the process to ensure that the stability requirements can be met.

In order to maintain a nearly equilateral-triangle formation for 5-year duration, one approach is to design orbits such that the three spacecraft acquire the same mean values of the semi-major axes, inclinations, and longitudes of ascending nodes, and meanwhile to keep the mean eccentricities as small as possible (cf. \ref{app:ecc}). Our steps for optimization are described as follows. 

\emph{Step 1}: For each spacecraft, we use the iterative relation (see \ref{app:eq_rv} for derivation)
\begin{equation} \label{eq_rv}
\bi{r}'_{0} = \left ( 1+\frac{1+\varepsilon }{1+4 \, \varepsilon } \frac{{\bar{a }}'-\bar{a }}{\bar{a }} \right )\bi{r}_{0}, \qquad 
\bi{v}'_{0} = \left ( 1- \frac{1+\varepsilon }{1+4 \, \varepsilon } \frac{{\bar{a }}'-\bar{a}}{2\bar{a }} \right )\bi{v}_{0},
\end{equation}
to have the average semi-major axis approaching the desired value $\bar{a}'=10^5$ km. Here ($\bi{r}_0$, $\bi{v}_0$) are the initial position and velocity before adjustment, and ($ \bi{r}'_{0} $, $ \bi{v}'_{0} $) the new ones. The value of $\bar{a}$ is determined by averaging the semi-major axis over the entire orbit generated from ($\bi{r}_0$, $\bi{v}_0$). Additionally, we define the parameter $ \varepsilon = (\bar{a}-a_{0})/a_{0} $. By applying the relation \eref{eq_rv} repeatedly, one can eliminate the long-term linear drift in arm lengths and breathing angles. 

Furthermore, we use the iterations (see \ref{app:eq_io} for derivation)
\begin{eqnarray} \label{eq_io}
i _{0}' = \left ( 1+ \frac{1+\epsilon }{1+4 \, \epsilon } \frac{\bar{i }' - \bar{i }}{\bar{i }} \right )i _{0}, \qquad 
\Omega_{0}^{'} = \Omega_{0} + \left({\bar{\Omega}}' -\bar{\Omega} \right),
\end{eqnarray}
to set the three spacecraft on the same average orbital plane. The primed and un-primed notation above is interpreted similarly as in equation \eref{eq_rv}, and $ \epsilon = (\bar{i}-i_{0})/i_{0} $.

\emph{Step 2}: To further improve the result from the previous step, we use numerical optimization to minimize the following cost function $C\!F_{12}$ \cite{Hughes2002}:
\begin{eqnarray}
C\!F_{12}=k_{1} C\! F_{1}+k_{2} C\!F_{2}, \label{33} \\
\eqalign{
C\!F_{1} \equiv \frac{1}{c_{1}}\int_{t_0}^{t_{f}}\left ( \left | v_{12} \right |+\left |v_{13}  \right |+\left | v_{23}  \right | \right ) \rmd t, \cr
C\!F_{2} \equiv \frac{1}{c_{2}}\int_{t_0}^{t_{f}}\left ( (\alpha _{1}-60^{\circ})^{2}+(\alpha _{2}-60^{\circ})^{2}+(\alpha _{3}-60^{\circ})^{2} \right ) \rmd t,} \label{312} 
\end{eqnarray} 
with weights $ k_{1} $ and $ k_{2} $ (typically, $0.5$) and normalization constants $c_1$ and $c_2$. The form of $C\!F_{12}$ is to bring down the relative velocities $|v_{ij}|$ and confine the breathing angles $\alpha_k$ close to $60^\circ$. The imposed constraints are directly taken from the stability requirements, i.e., $ \left |v_{ij}  \right | \leqslant 5\,\text{m/s} $ and $ \left |\alpha _{k}  \right |\leqslant 0.1^{\circ} $ for the first two years, and loosened to $ \left |v_{ij}  \right | \leqslant 10\,\text{m/s} $ and $ \left |\alpha _{k}  \right |\leqslant 0.2^{\circ} $ for the following three years. The independent variables one may vary include the initial eccentricities, arguments of periapsis, and true anomalies of the three spacecraft. 

Minimizing the cost function $C\!F_{12}$ can help reduce the average eccentricities, which, as our simulations have shown, strongly affect the formation stability (cf. \ref{app:ecc}) in the long run. To achieve better results in numerical search, the step 2 may be repeated. 

\emph{Step 3}: Redo the iteration \eref{eq_rv} from the first step if the semi-major axes stray from $ 10^{5} $ km after the step 2.

  
\section{Optimized orbits for TianQin}

\begin{figure}[!ht]
\centering 
\begin{minipage}{3in}
\includegraphics[width=3in,height=2.02125in]{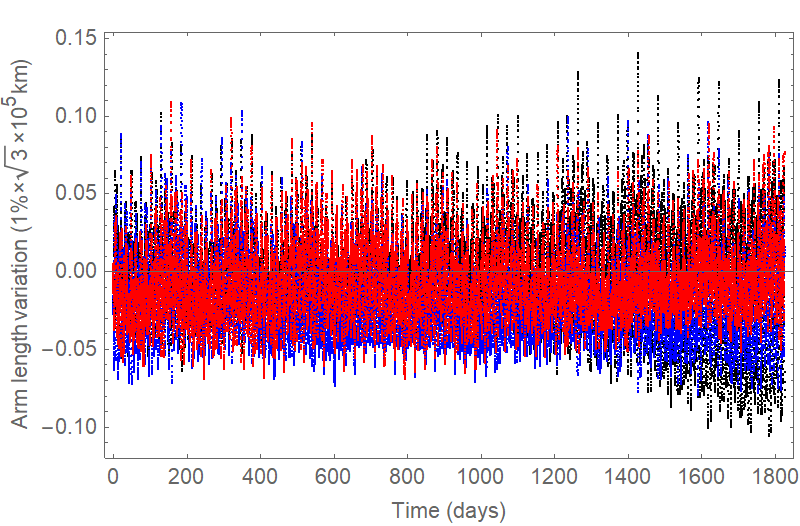}
\end{minipage}
\begin{minipage}{3in}
\includegraphics[width=3in,height=2.02125in]{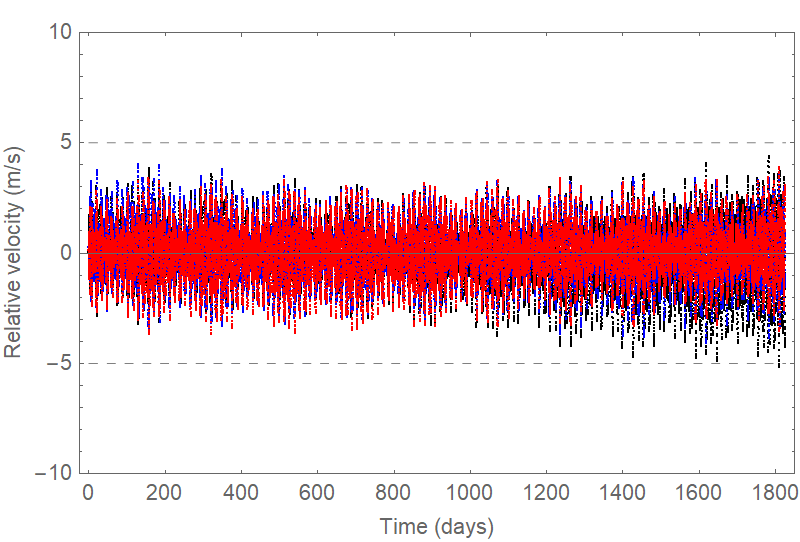}
\end{minipage}

\begin{minipage}{3in}
\includegraphics[width=3in,height=2.02125in]{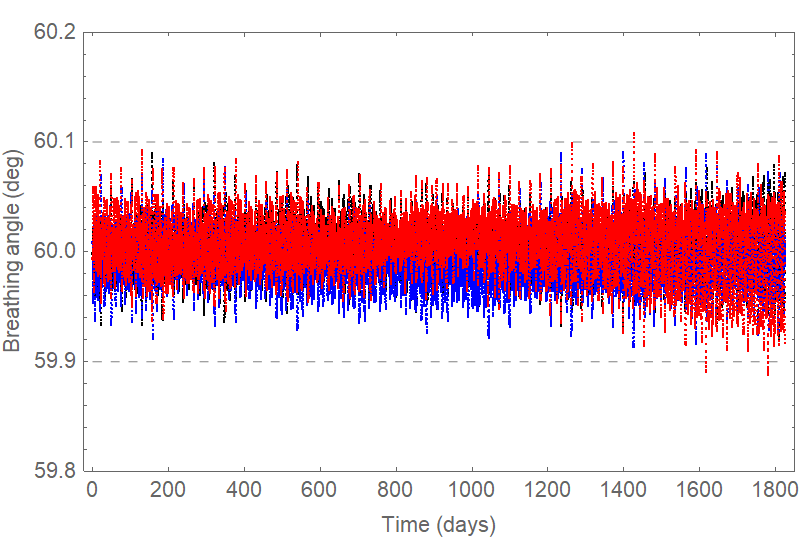}
\end{minipage}
\begin{minipage}{3in}
\includegraphics[width=3in,height=2.02125in]{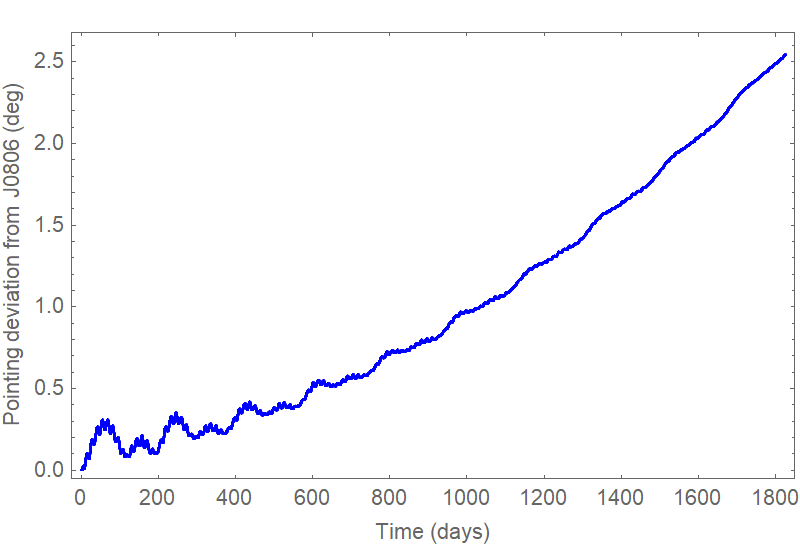}
\end{minipage}
\caption{\label{fig_tq} Evolution of constellation geometry of the optimized TianQin orbits generated from the initial elements of table \ref{tab_tq}. The plots show 5-year variations of the arm lengths $L_{12}$ (black), $L_{13}$ (blue), $L_{23}$ (red), and the relative velocities $v_{12}$ (black), $v_{13} $ (blue), $v_{23}$ (red), and the breathing angles $\alpha_{1}$ (black), $\alpha_{2}$ (blue), $\alpha_{3}$ (red), and the pointing deviation from J0806. } 
\end{figure}

The initial orbital elements obtained from optimization are listed in table \ref{tab_tq}. We present, in figure \ref{fig_tq}, the 5-year evolutions of the arm lengths $L_{12}$ (black), $L_{13}$ (blue), $L_{23}$ (red), and the relative velocities $v_{12}$ (black), $v_{13} $ (blue), $v_{23}$ (red), and the breathing angles $\alpha_{1}$ (black), $\alpha_{2}$ (blue), $\alpha_{3}$ (red), as well as the pointing deviation $\phi$ from J0806 (the angle between the direction to J0806 and the normal direction of the triangle). The result has been independently verified by other orbit simulators. For more details, we summarize the stability performance in table \ref{tab_all}. One can see that the optimized TianQin orbits fulfill the stability requirements provided in table \ref{tab_req}. 

\begin{table}[!ht]
\caption{\label{tab_tq} The initial elements of the optimized TianQin orbits in the EarthMJ2000Ec (Keplerian, ecliptic) and EarthMJ2000Eq (Cartesian, equatorial) coordinates at the epoch 22 May, 2034 12:00:00 UTC. The subsequent orbital evolution is shown in figure \ref{fig_tq}. }
\lineup
\footnotesize\rm
\begin{tabular*}{\textwidth}{@{}l*{15}{@{\extracolsep{0pt plus12pt}}l}}
\br
 & \0$  a $\,(km) &  $  e $ & $ i $\,($ ^{\circ} $) & $ \Omega $\,($ ^{\circ} $) & $  \omega   $\,($ ^{\circ} $)  & \0$ \nu $\,($ ^{\circ} $) \\ 
\mr
SC1&  \099\,995.572\,323  & 0.000\,430  & 94.697\,997 & 210.445\,892 &   358.624\,463  &  \061.329\,603 \\
SC2&   100\,011.400\,095  & 0.000\,000  & 94.704\,363 & 210.440\,199 & \0\00.000\,000  &  179.930\,706 \\
SC3&  \099\,993.041\,899  & 0.000\,306  & 94.709\,747 & 210.444\,582 & \0\00.001\,624  & 299.912\,164  \\
\br
\end{tabular*}
\lineup
\footnotesize\rm
\begin{tabular*}{\textwidth}{@{}l*{15}{@{\extracolsep{0pt plus12pt}}l}}
\br
 & $  x $\,(km) &  $  y $\,(km) & $ z $\,(km) & $ v_{x} $\,(km/s) & $  v_{y}   $\,(km/s)  & $ v_{z} $\,(km/s) \\ 
\mr
SC1& \-46\,746.087\,307  &  \-51\,973.844\,583 &  71\,473.835\,818  &  1.448\,401 &   0.471\,646 & 1.291\,321\\
SC2&  86\,220.582\,041 & 46\,448.360\,669  & 20\,269.217\,366  &  0.085\,035 &  0.663\,048 & \-1.881\,140 \\
SC3& \-39\,378.654\,985  & \0\,5547.379\,475  & \-91\,728.424\,823  & \-1.533\,416  &   \-1.134\,792 &0.590\,239 \\
\br
\end{tabular*}
\end{table}

\begin{table}[!ht]
\caption{\label{tab_all}A summary of the optimization results regarding their average orbital planes over 5 years and stability performance of the constellation.}
\lineup
\footnotesize
\begin{tabular}{@{}lllllll}
\br
&\centre{2}{Orbital plane}&\centre{4}{Stability for 5 years (and the first 2 years)}\\
\ns
Result&\crule{2}&\crule{4}\\
 &$ \bar{\Omega} $\,($ ^{\circ} $)&\0$ \bar{i} $\,($ ^{\circ} $)&$ \left | \Delta l_{ij} \right |_{\max} $($ \% $)&$ \left | v_{ij} \right |_{\max} $($ \text{m/s} $)&$ \left |\Delta \alpha _{k}\right |_{\max}  $($ ^{\circ} $)&\0$  \bar{\phi}_{ -(\bar{\phi}-\phi_{\min})}^{+(\phi_{\max}-\bar{\phi})} $\,($ ^{\circ} $)\\
\mr
TianQin$^{\rm a}$ &211.42&\094.62&0.140 (0.109)&5.178 (4.003)&0.112 (0.092)&\0$ 1.00_{-1.00}^{+1.54} $ ($ 0.32_{-0.31}^{+0.27} $)\\
P1 &210.18&\091.63&0.156 (0.098)&4.993 (4.130)&0.150 (0.090)&\0$ 3.08_{-0.28}^{+0.29} $ ($ 3.10_{-0.26}^{+0.27} $)\\
P2 &240.00&\088.98&0.151 (0.125)&5.260 (4.626)&0.139 (0.098)&$ 30.08_{-0.24}^{+0.61} $ ($ 30.28_{-0.32}^{+0.40} $)\\
P3 &270.00&\086.58&0.164 (0.126)&6.005 (4.793)&0.160 (0.102)&$ 60.05_{-0.34}^{+0.75} $ ($ 60.38_{-0.37}^{+0.43} $)\\
P4 &120.00&100.00&0.148 (0.131)&5.423 (4.319)&0.132 (0.102)&$ 88.77_{-1.54}^{+1.23} $ ($ 88.94_{-1.07}^{+1.06} $)\\
P5$^{\rm b}$ &330.00&\090.00&0.161 (0.119)&5.773 (4.458)&0.142 (0.093)&$ 60.55_{-1.14}^{+1.28} $ ($ 61.32_{-0.53}^{+0.51} $)\\
P6$^{\rm c}$ &180.00&\090.00&0.136 (0.091)&5.333 (4.167)&0.120 (0.083)&$ 30.77_{-1.20}^{+0.86} $ ($ 30.11_{-0.53}^{+0.58} $)\\
\br
\end{tabular}\\
$^{\rm a}$ Initial epoch 22 May, 2034 12:00:00 UTC; for the others, 01 January, 2034 00:00:00 UTC. \\
$^{\rm b}$ Retrograde orbits. \\
$^{\rm c}$ The detector pointing deviates from the Galactic Center by $ 6.45_{-0.59}^{+0.78} $ ($ 6.77_{-0.50}^{+0.46} $) degree.
\end{table}
\normalsize


\section{Secular pointing drift} \label{sec:Change}

In figure \ref{fig_tq} (the lower right panel), one observes an overall increase in the deviation angle $\phi$ away from J0806. In what follows, we show that the pointing shift is mainly caused by orbital precession under the combined influence of lunisolar and the Earth's $J_2$ oblateness perturbations. 

\begin{figure}[!ht]
\centering 
\begin{minipage}{3in}
\includegraphics[width=3in,height=2.02125in]{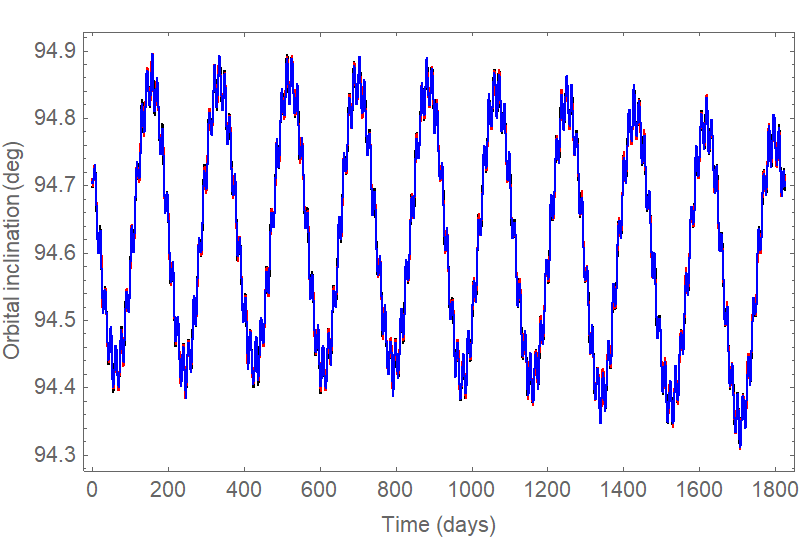}
\end{minipage}
\begin{minipage}{3in}
\includegraphics[width=3in,height=2.02125in]{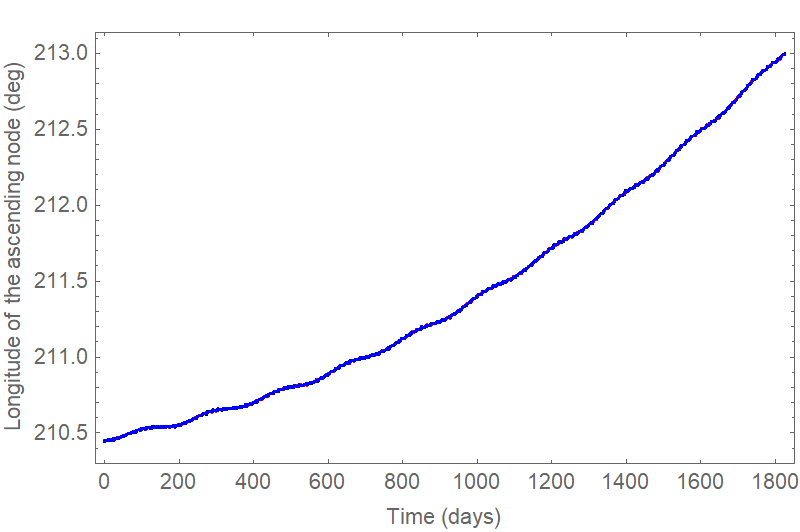}
\end{minipage}
\caption{\label{fig_io} Evolution of the inclinations and longitudes of the ascending nodes of the three spacecraft (marked black, blue, and red), for the optimized TianQin orbits of table \ref{tab_tq}. Both curves appear blue due to extensive overlap. } 
\end{figure}

In terms of the elements $i$ and $\Omega$, the unit normal vector $\hat{\bi{n}}$ of an orbital plane can be expressed as
\begin{eqnarray}
\hat{\bi{n}} = \begin{pmatrix}
\sin  i \sin \Omega \\ 
-\sin  i \cos \Omega \\ 
\cos i \end{pmatrix}. \label{34}
\end{eqnarray}
Thus one can calculate the shift angle from $\hat{\bi{n}}_{0}(i_0,\Omega_0) $ to $ \hat{\bi{n}}_{t}(i(t),\Omega(t)) $ according to $\cos\phi(t)=\hat{\bi{n}}_{t} \cdot \hat{\bi{n}}_{0}$ with $\hat{\bi{n}}_{0}$ pointing at J0806. Let $\Omega(t)=\Omega_{0}+\Delta \Omega(t)$ and $i(t)=i_{0}+\Delta i(t)$. Taking into account that both $\Delta i$ and $\Delta \Omega$ are small, we have an approximation
\begin{eqnarray}
\phi^2 \approx \Delta i^2 + \Delta \Omega^2 \sin^2 i_0. \label{37}
\end{eqnarray}
As figure \ref{fig_io} shows, $ \Delta \Omega $ outgrows $ \Delta i$ and $ \sin i_0 \sim 1$. Therefore the deviation angle $\phi$ is dominated by $ \Delta \Omega $ in the long run ($ \left |\Delta \Omega \right |_{\max}= 2.55^{\circ} $, $ \left |\Delta i \right |_{\max}= 0.40^{\circ} $), which manifests as a common trend in both figure \ref{fig_tq} (lower right) and figure \ref{fig_io} (right). One can understand the behavior of $ \Delta \Omega $ and $ \Delta i$ as a generic property that $ \Omega $ possesses secular change, while not so much for $ i $ \cite{Smith1962}. Our numerical tests over $ 85^{\circ} \leq i_{0} \leq 95^{\circ} $ also confirm this observation.

From perturbative analysis, the mean rate of secular change of $ \Omega $ is given by \cite{Smith1962} (also \cite{Vallado1997}, see p 614) 
\begin{eqnarray}
\dot{\Omega}_{m/s} = -\frac{3 G M_{m/s} (1-e^{2})^{3}(1+5e^{2})}
{4 a_{m/s}^3 n (1-e_{m/s}^2)^{3/2}} \cos i_{mo} \label{39}
\end{eqnarray}
for the third-body perturbation from the Moon ($ m $) or the Sun ($ s $). Here $ M_{m/s} $ denotes the mass of the third body, and $n$ the mean angular motion of the spacecraft. All the orbital elements, e.g., $e$ and $a_{m/s}$, take on mean values over certain periods (e.g., one lunar month for the Moon). Note that $ i_{mo} $ represents the mean inclination to the third body's orbital plane. Likewise for the Earth's $ J_{2} $ perturbation, one has \cite{Liu2000} (see p 139)
\begin{eqnarray}
\dot{\Omega }_{J_{2}}=-\frac{3J_{2}R_{\oplus }^{2}n}{2[a(1-e^{2})]^{2}}\cos i_{eq}, \label{40}
\end{eqnarray}
where $i_{eq}$ denotes the mean inclination to the Earth's equator, and $ R_{\oplus} $ the Earth's equatorial radius and $ J_{2}=1.08263\times 10^{-3} $. The equation is commonly found in the context of Sun-synchronous orbits. 

The formulas \eref{39} and \eref{40} give rise to a total secular change of $ 2.36^{\circ} $ in $\Omega$, which combines the contributions from the Moon ($+ 1.60^{\circ} $), the Sun ($+1.08^{\circ} $), and the $J_2$ oblateness ($ -0.32^{\circ} $). The value agrees with the numerical result $ \left |\Delta \Omega \right |_{\max} = 2.55^\circ$ within $8\%$. It confirms that the long-term pointing shift $ \phi $ in figure \ref{fig_tq} is primarily driven by lunisolar gravitational perturbations. 

One can see from \eref{39} and \eref{40} that there will be no orbital precession if $ i_{mo/eq} = 90^{\circ} $. However, the equality cannot be achieved simultaneously for the three perturbing bodies. Because the Moon's effect account for the largest one, we can arrange the mean orbital plane perpendicular to the mean lunar orbital plane to reduce $ \left |\Delta \Omega \right |_{\max} $, and consequently, $ \left |\Delta \phi \right |_{\max} $ as well. This is demonstrated in the next section (see the last plot of figure \ref{p1}).


\section{Optimized orbits for other pointings}

From the previous analysis, we have found that the orientation of the TianQin orbital plane varies about $2.5^\circ$ during a 5-year mission lifetime, indicating a rather stable detector pointing at J0806. Considering the future possibilities of new reference sources, we present the optimization results for other pointing directions, in this section. 

\begin{figure}
\centering 
\begin{minipage}{3in}
\includegraphics[width=3in,height=2.02125in]{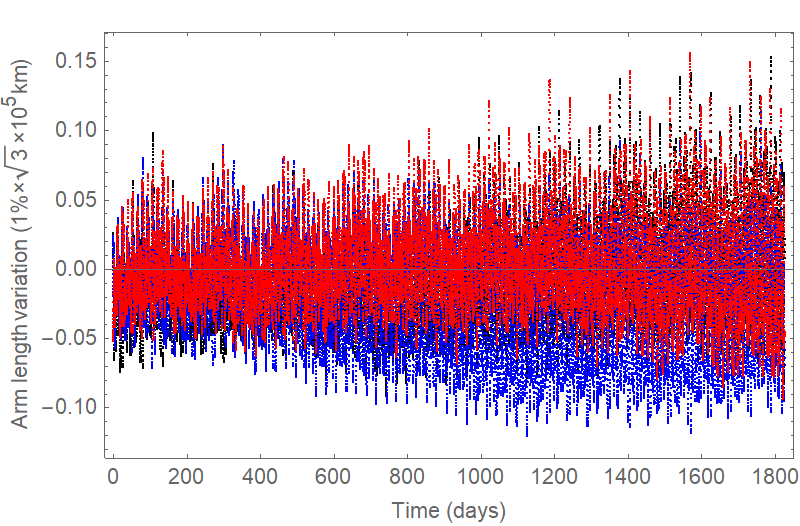}
\end{minipage}
\begin{minipage}{3in}
\includegraphics[width=3in,height=2.02125in]{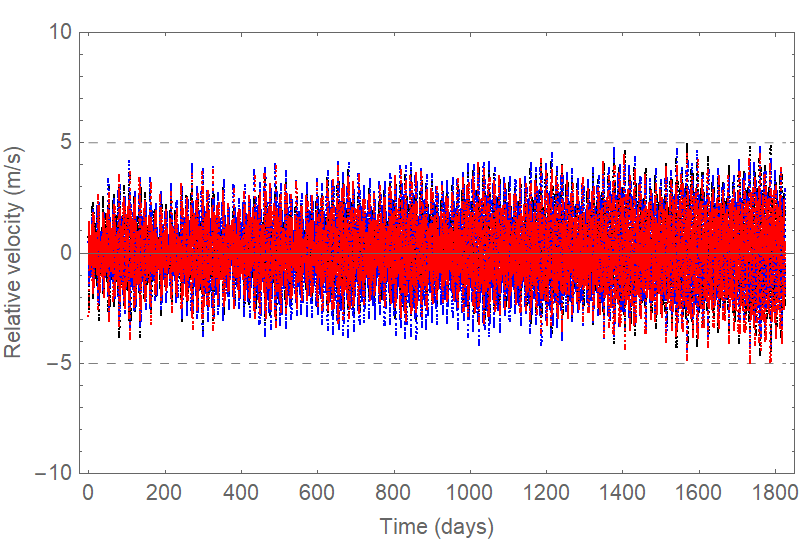}
\end{minipage}
\begin{minipage}{3in}
\includegraphics[width=3in,height=2.02125in]{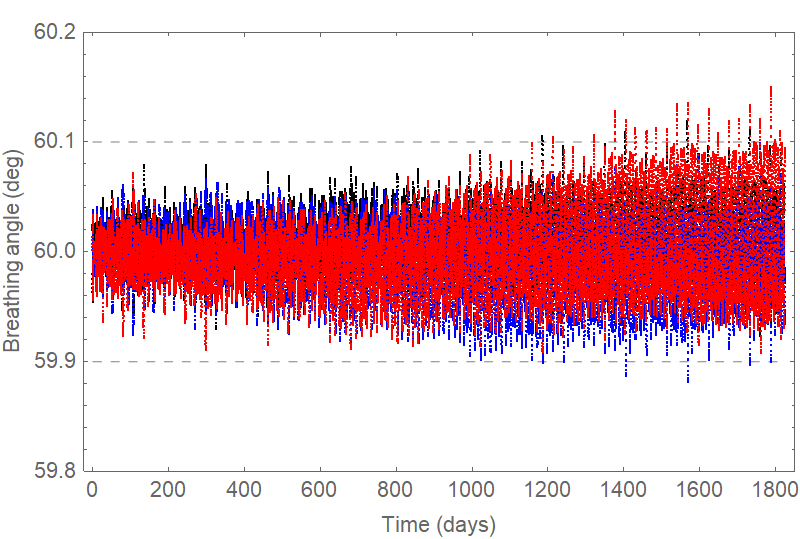}
\end{minipage}
\begin{minipage}{3in}
\includegraphics[width=3in,height=2.02125in]{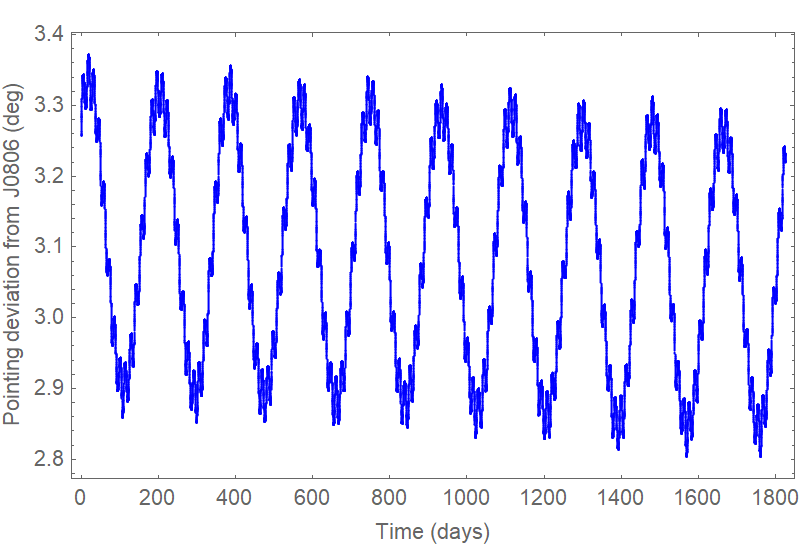}
\end{minipage}
\caption{Constellation evolution of the optimized orbits P1 (see figure \ref{fig_tq} for the color assignment). } 
\label{p1}
\end{figure}

\begin{figure}
\centering 
\begin{minipage}{3in}
\includegraphics[width=3in,height=2.02125in]{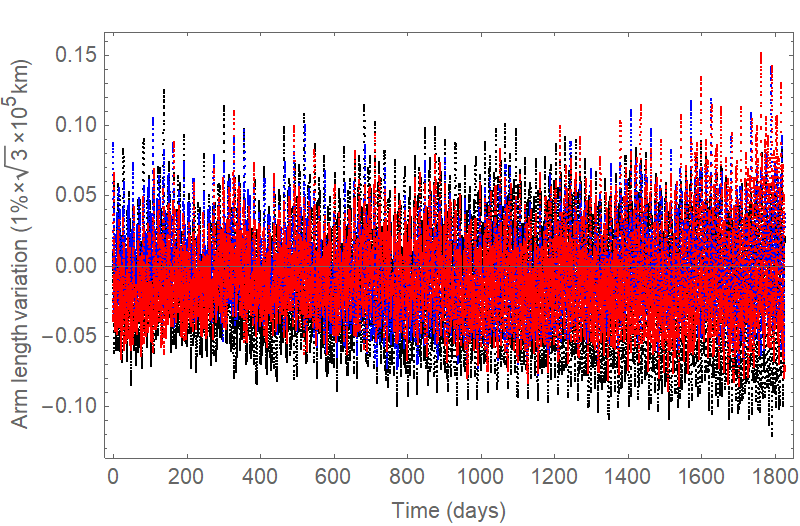}
\end{minipage}
\begin{minipage}{3in}
\includegraphics[width=3in,height=2.02125in]{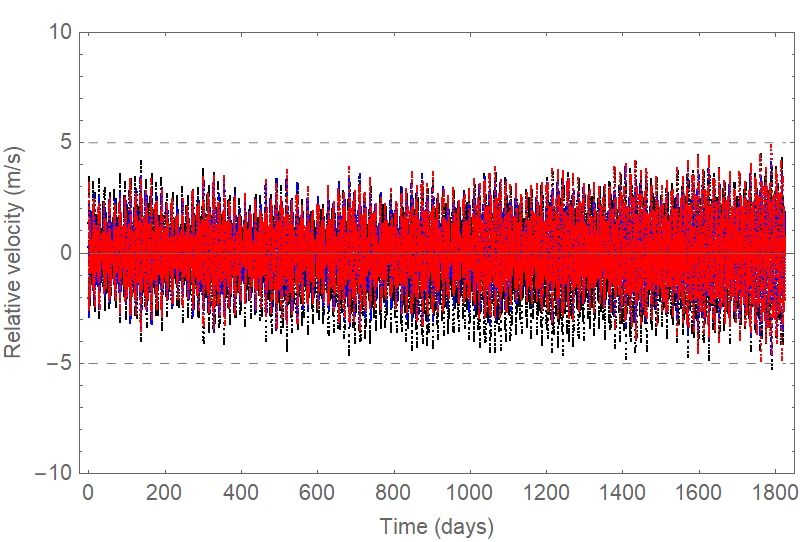}
\end{minipage}
\begin{minipage}{3in}
\includegraphics[width=3in,height=2.02125in]{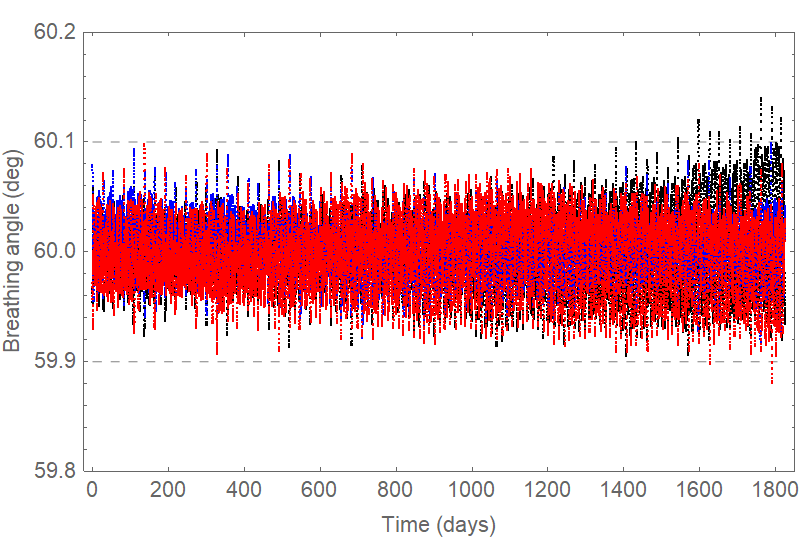}
\end{minipage}
\begin{minipage}{3in}
\includegraphics[width=3in,height=2.02125in]{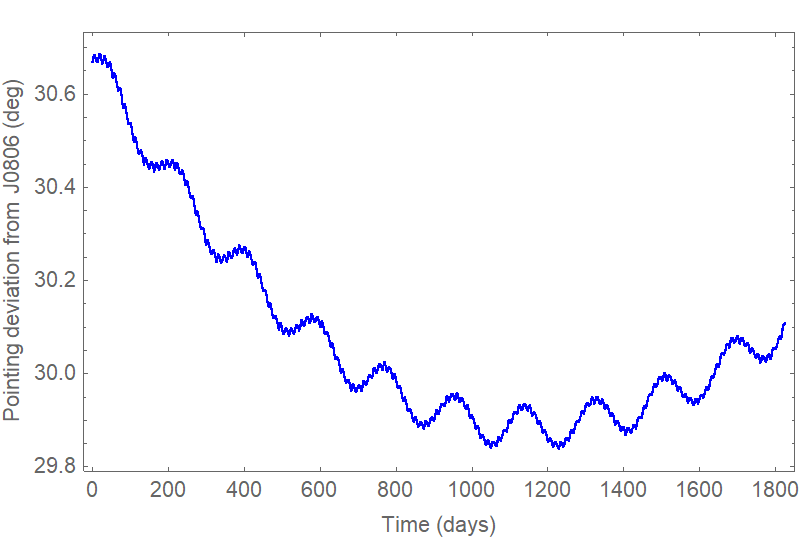}
\end{minipage}
\caption{Constellation evolution of the optimized orbits P2 (see figure \ref{fig_tq} for the color assignment). } 
\label{p2}
\end{figure}

\begin{figure}
\centering 
\begin{minipage}{3in}
\includegraphics[width=3in,height=2.02125in]{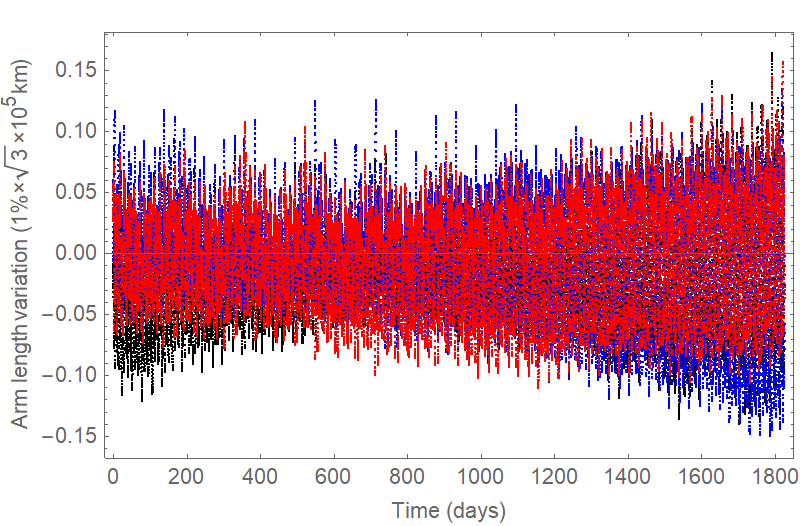}
\end{minipage}
\begin{minipage}{3in}
\includegraphics[width=3in,height=2.02125in]{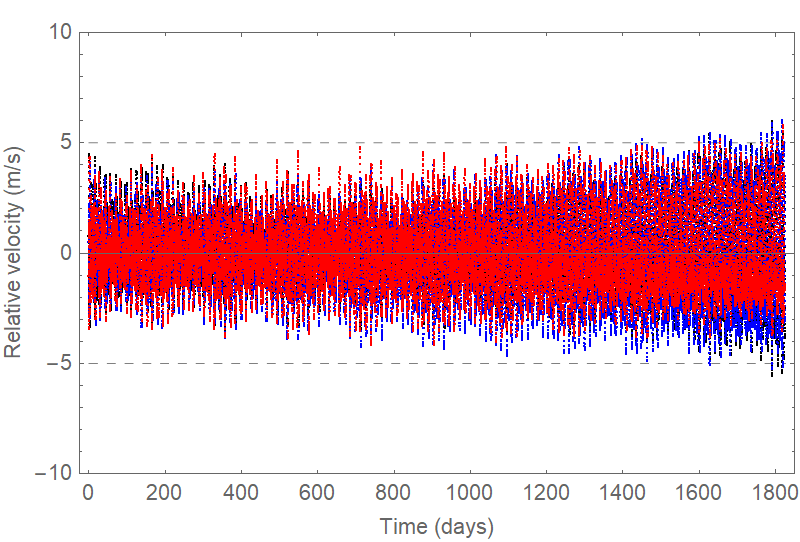}
\end{minipage}
\begin{minipage}{3in}
\includegraphics[width=3in,height=2.02125in]{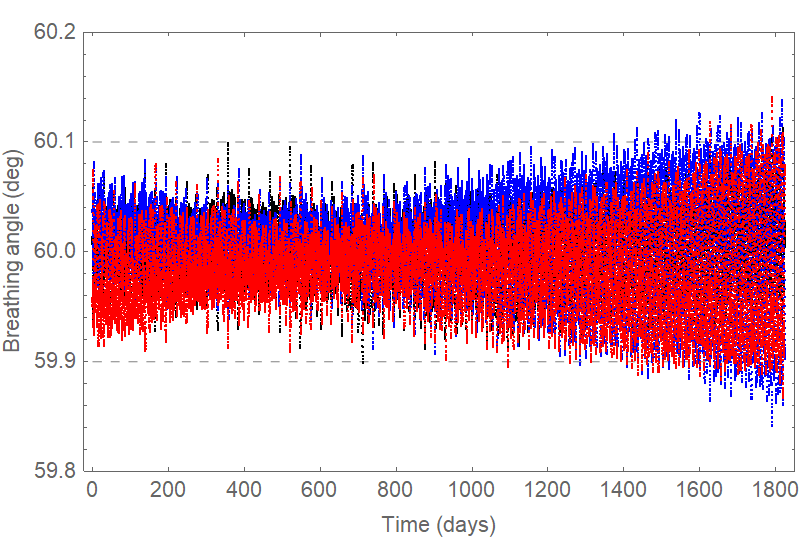}
\end{minipage}
\begin{minipage}{3in}
\includegraphics[width=3in,height=2.02125in]{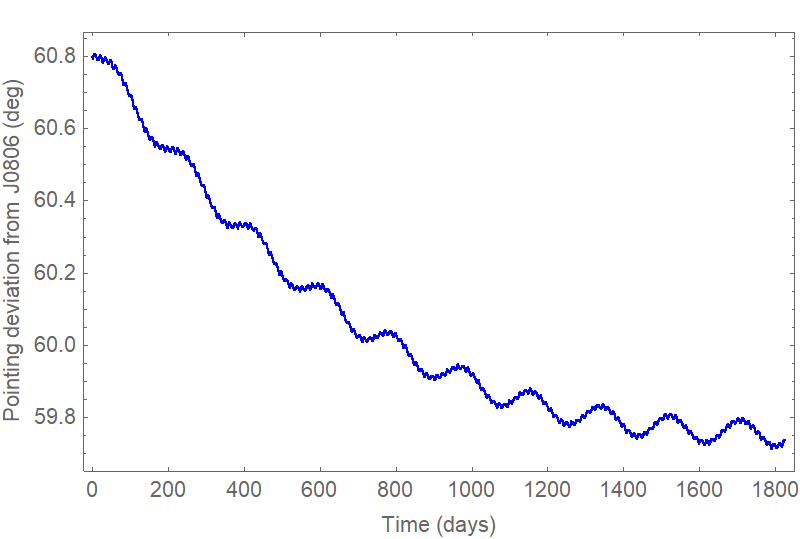}
\end{minipage}
\caption{Constellation evolution of the optimized orbits P3 (see figure \ref{fig_tq} for the color assignment). } 
\label{p3}
\end{figure}

\begin{figure}
\centering 
\begin{minipage}{3in}
\includegraphics[width=3in,height=2.02125in]{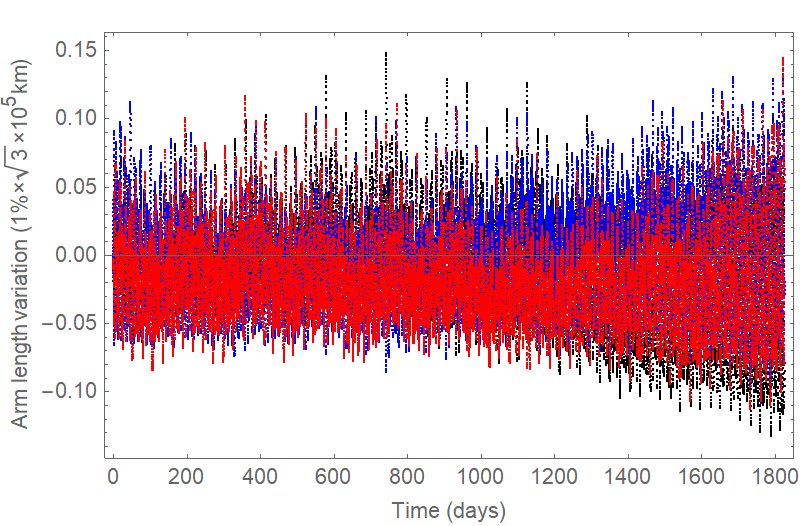}
\end{minipage}
\begin{minipage}{3in}
\includegraphics[width=3in,height=2.02125in]{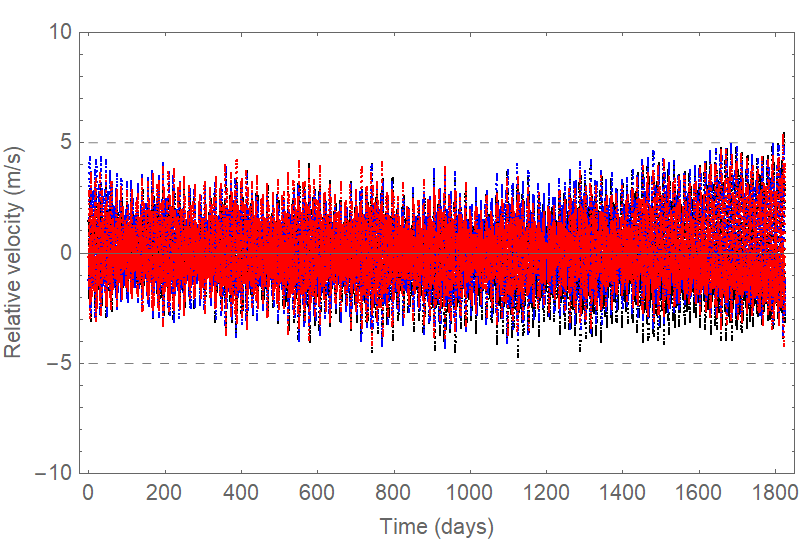}
\end{minipage}
\begin{minipage}{3in}
\includegraphics[width=3in,height=2.02125in]{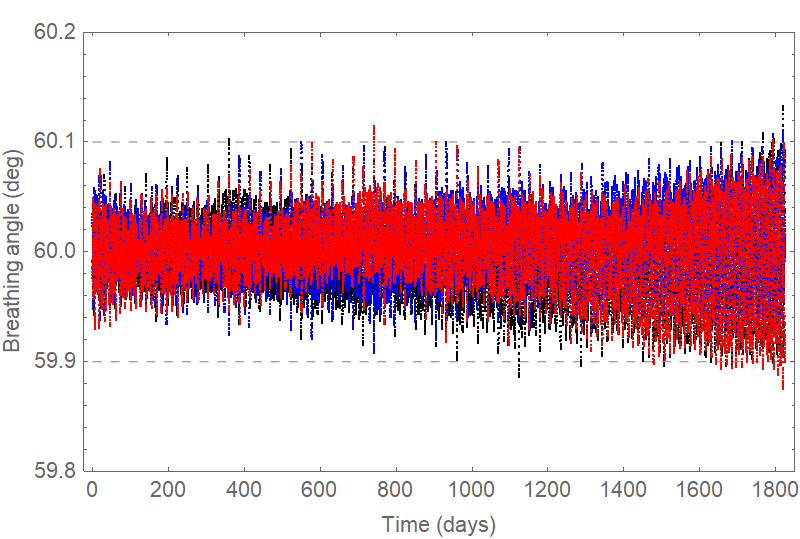}
\end{minipage}
\begin{minipage}{3in}
\includegraphics[width=3in,height=2.02125in]{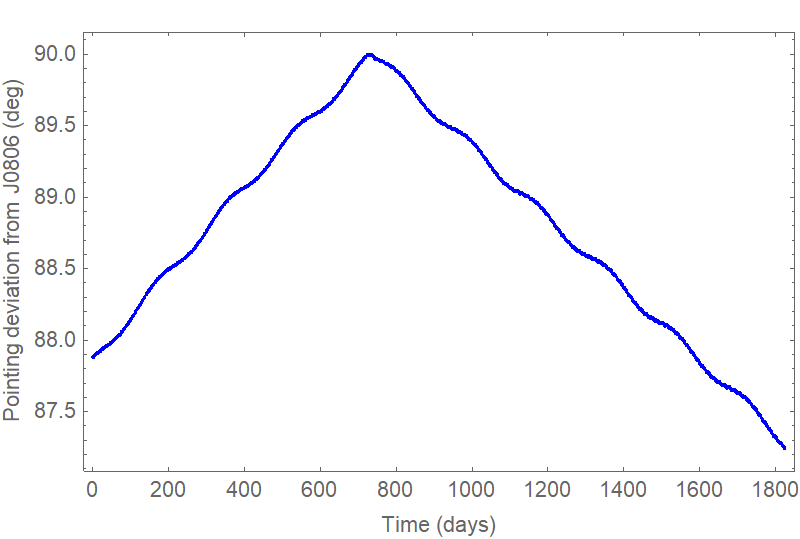}
\end{minipage}
\caption{Constellation evolution of the optimized orbits P4 (see figure \ref{fig_tq} for the color assignment). We request the deviation angle $\leq 90^\circ$, hence the broken line in the lower right plot. } 
\label{p4}
\end{figure}

\begin{figure}
\centering 
\begin{minipage}{3in}
\includegraphics[width=3in,height=2.02125in]{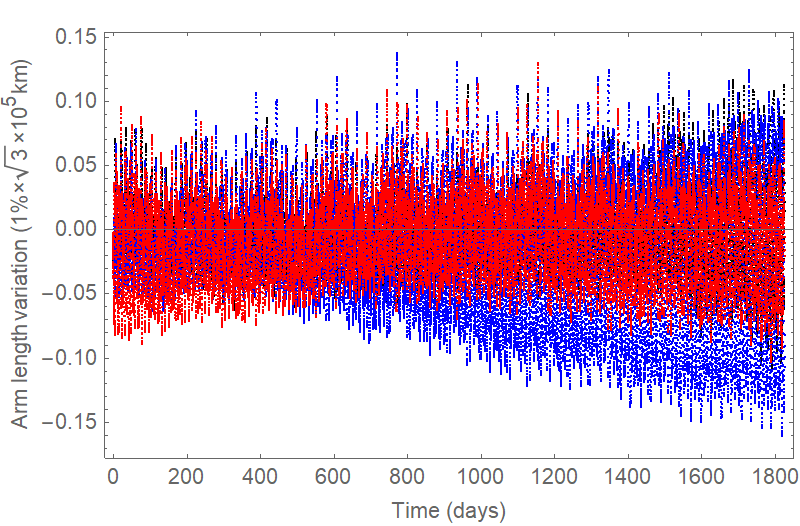}
\end{minipage}
\begin{minipage}{3in}
\includegraphics[width=3in,height=2.02125in]{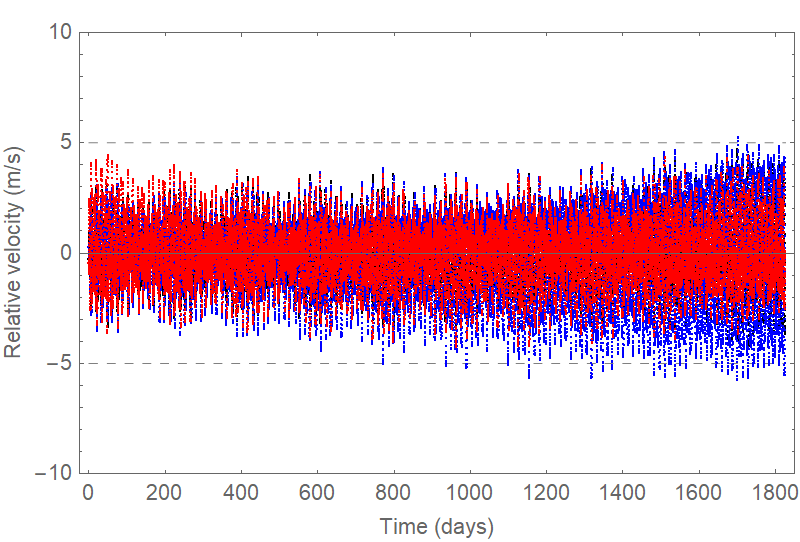}
\end{minipage}
\begin{minipage}{3in}
\includegraphics[width=3in,height=2.02125in]{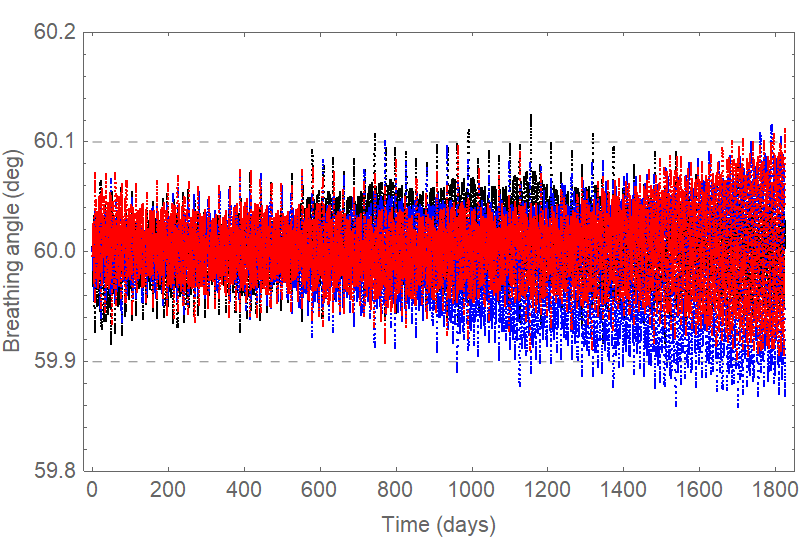}
\end{minipage}
\begin{minipage}{3in}
\includegraphics[width=3in,height=2.02125in]{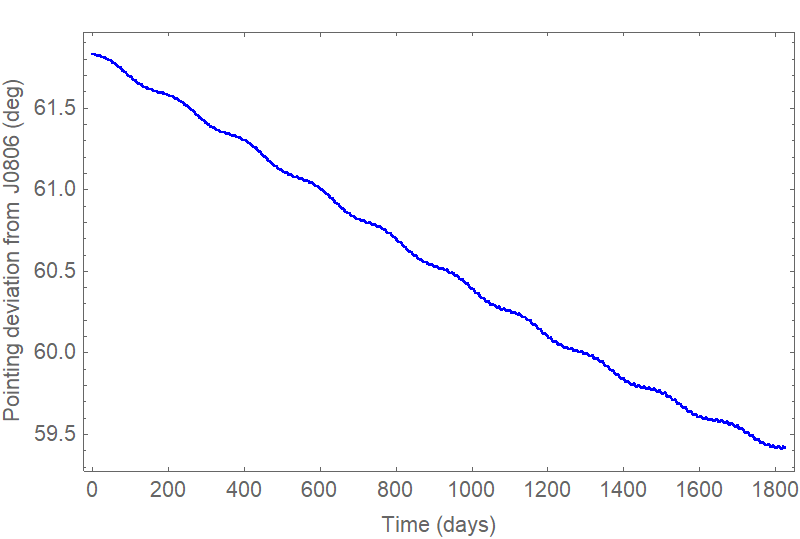}
\end{minipage}
\caption{Constellation evolution of the optimized orbits P5 (see figure \ref{fig_tq} for the color assignment). } 
\label{p5}
\end{figure}

\begin{figure}
\centering 
\begin{minipage}{3in}
\includegraphics[width=3in,height=2.02125in]{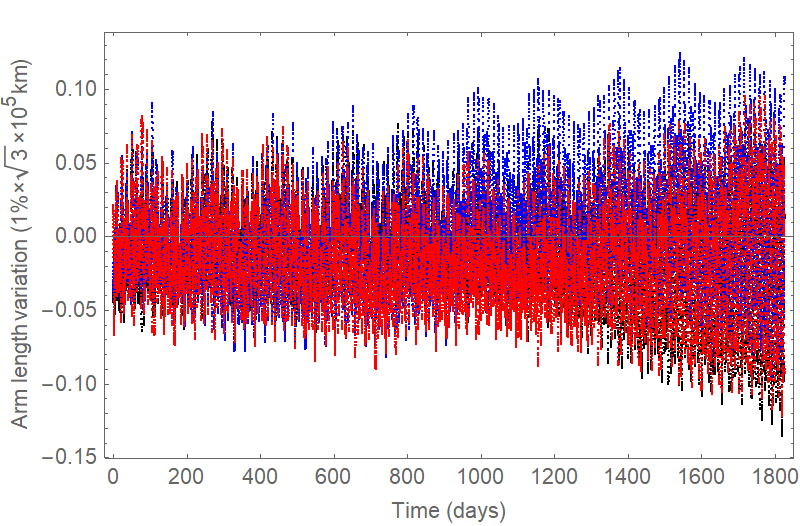}
\end{minipage}
\begin{minipage}{3in}
\includegraphics[width=3in,height=2.02125in]{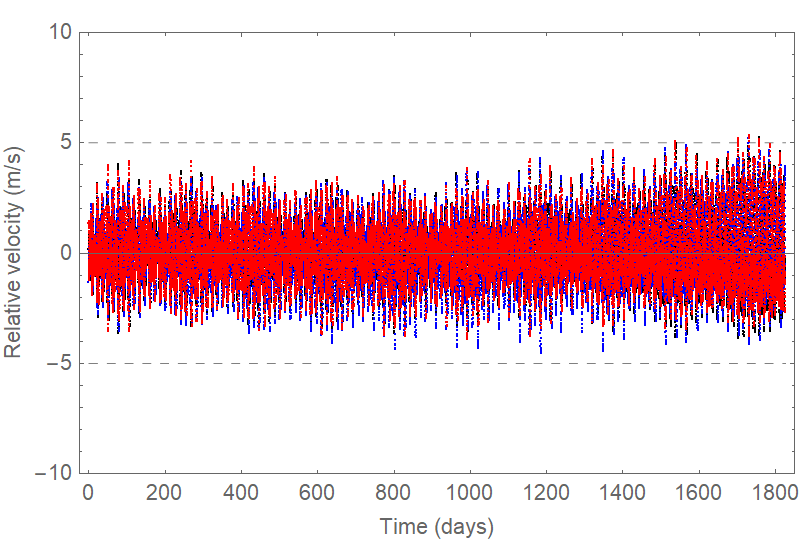}
\end{minipage}
\begin{minipage}{3in}
\includegraphics[width=3in,height=2.02125in]{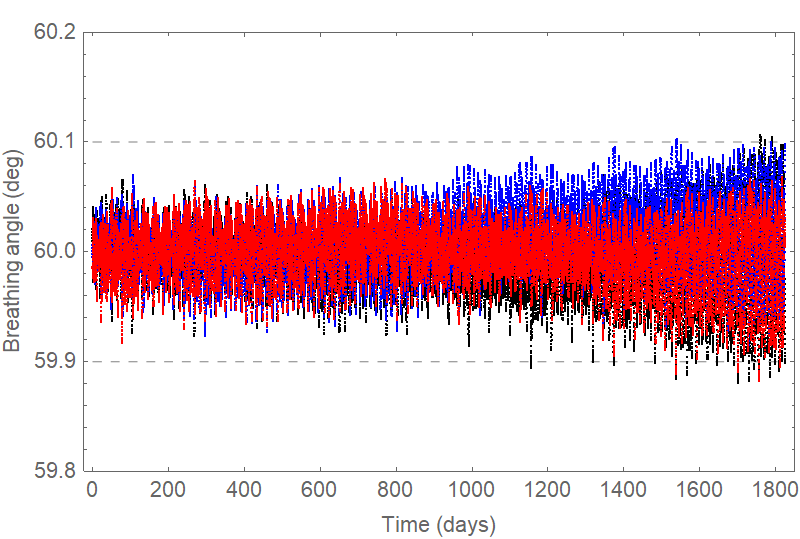}
\end{minipage}
\begin{minipage}{3in}
\includegraphics[width=3in,height=2.02125in]{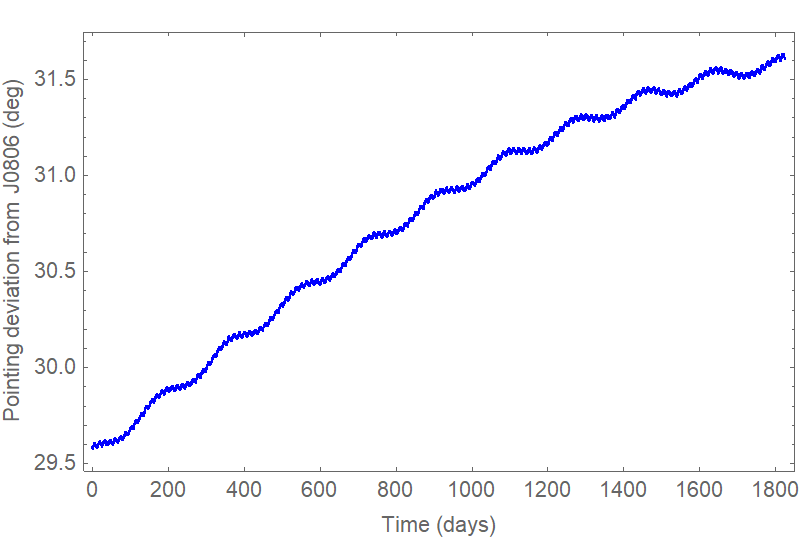}
\end{minipage}
\caption{Constellation evolution of the optimized orbits P6 (see figure \ref{fig_tq} for the color assignment). } 
\label{p6}
\end{figure}

As mentioned before, the long-term stability of the constellation depends mainly on the magnitude of the average eccentricities of the orbits. From our numerical tests, we have noticed that the average eccentricity can attain small values if one sets the orbital plane roughly perpendicular to the mean orbital plane of the Moon. Therefore to ease our search in optimization, we consider six detector pointing directions P1-6, all approximately aligned with the lunar orbital plane, and $30^{\circ}$ apart to spread over a half circle. Regarding the mean lunar orbital plane (from 1 January, 2034 to 1 January, 2039) in the EarthMJ2000Ec coordinates, it has the inclination $ 5.161\,139^{\circ} $ and the longitude of ascending node $138.590\,584^{\circ}$ \cite{JPL}. Our results are shown in table \ref{tab_p16} and figures \ref{p1}-\ref{p6}. Their stability performance is summarized in table \ref{tab_all}. 

For P1, we point out that the orbital configuration resembles TianQin's (see table \ref{tab_all}). The main difference lies in that the long-term growth of the deviation angle (figure \ref{fig_tq}) is suppressed in P1, and taken over by semi-annual fluctuation of $0.5^\circ$ due to the inclination (figure \ref{p1}).

In the cases of P3 and P4 (figures \ref{p3}, \ref{p4}), the time evolution of the breathing angles can wander over $\pm 0.1^{\circ}$ in the first two years, but only at a few peaks and by a small amount ($0.002^{\circ}$, see table \ref{tab_all}). Hence we still include them for future consideration. Unfortunately in these two cases, further improvement to suppress the angle excursion within $\pm 0.1^{\circ}$ appears difficult and time-consuming. 

Regarding P6 (figure \ref{p6}), it is worth noting that the orbital pointing differs from the direction to the Galactic Center (the compact astronomical radio source Sagittarius A*) for only about $ 6.5^{\circ} $ (see table \ref{tab_all}), the smallest among all the cases. 

\begin{table}
\caption{\label{tab_p16} The initial elements of the optimized orbits with the pointings P1-6 in the EarthMJ2000Ec coordinates and at the epoch 01 January 2034 00:00:00 UTC. Their time evolutions correspond to figures \ref{p1}-\ref{p6}, respectively. }
\lineup
\footnotesize\rm
\begin{tabular*}{\textwidth}{@{}l*{15}{@{\extracolsep{0pt plus12pt}}l}}
\br
Ptn & & \0$a$\,(km) &  $e$ & \0$i$\,($^\circ$) & $\Omega$\,($^\circ$) & \0\0$\omega$\,($^\circ$)  & \0$\nu$\,($^\circ$) \\ 
\mr
P1&SC1 &\099\,988.451\,891   &0.000\,000  &\091.447\,831   &210.436\,268  &\0\00.000\,000   &\060.019\,828   \\
&SC2&100\,047.700\,990   &0.000\,804  &\091.444\,569 &210.444\,886  &180.451\,960   &359.620\,218   \\
&SC3&\099\,985.159\,485   &0.000\,829  &\091.445\,774   &210.436\,648 &\084.711\,013   &215.305\,422   \\
\mr
P2&SC1&\099\,985.313\,256   &0.000\,694  &\088.977\,641   &240.599\,044  &319.646\,851   &100.370\,180  \\
&SC2&100\,063.708\,616   &0.000\,544  &\088.974\,726   &240.598\,772  &179.927\,490   &\0\00.012\,642   \\
&SC3&\099\,975.723\,887   &0.000\,928  &\088.984\,674   &240.600\,376  &\057.832\,469   &242.028\,312  \\
\mr
P3&SC1&\099\,992.403\,653   &0.000\,644  &\086.746\,616   &270.774\,164  &314.121\,261   &103.509\,222   \\
&SC2&100\,033.833\,610   &0.000\,000  &\086.749\,894   &270.770\,211  &\0\00.312\,080   &177.221\,271   \\
&SC3&\099\,983.589\,342   &0.000\,703  &\086.746\,990   &270.776\,583  &\063.576\,742   &233.967\,848   \\
\mr
P4&SC1&\099\,984.187\,480   &0.000\,607  &100.086\,542   &117.446\,322  &289.515\,853   &174.493\,735   \\
&SC2&100\,007.078\,264   &0.000\,232  &100.082\,574   &117.432\,353  &226.666\,677   &357.274\,707   \\
&SC3&100\,008.968\,474   &0.000\,208  &100.084\,516   &117.444\,713  &\0\00.056\,315  &343.906\,747   \\
\mr
P5&SC1&\099\,993.147\,430   &0.000\,091  &\089.995\,647   &328.718\,323  &\0\00.024\,628   &\060.172\,294 \\
&SC2&100\,011.119\,344   &0.000\,274  &\089.989\,041   &328.724\,788  &234.731\,987   &305.458\,669   \\
&SC3&\099\,995.665\,243   &0.000\,000  &\089.984\,838   &328.717\,370  &\036.310\,671   &263.883\,611   \\
\mr
P6&SC1&100\,000.269\,197   &0.000\,003  &\089.751\,362   &181.247\,381  &131.469\,664   &180.050\,283   \\
&SC2&\099\,994.745\,911   &0.000\,027  &\089.759\,135   &181.250\,998  &359.603\,358   &\071.845\,840   \\
&SC3&100\,011.728\,446   &0.000\,769  &\089.754\,793   &181.247\,289  &190.317\,712   &\0\01.173\,907   \\
\br
\end{tabular*}
\end{table}

\section{Conclusion}

In order to achieve a successful mission, it is vital for the TianQin constellation to attain high stability in orbit design and operation. In this study, we apply optimization methods and manage to stabilize the TianQin constellation down to the level of $\pm 0.1\%$ in arm lengths, $\pm 4$ m/s in relative velocities, and $\pm 0.1^\circ$ in breathing angles, for a period of 2 years along free-fall orbits. The optimized orbit configuration fulfills, with margins, the assumed 5-year stability requirements from instrumentation. Though no direct constraint is imposed, the detector pointing can be made quite stable, but a small amount of pointing variation ($<2.6^\circ$ in 5 years) is unavoidable due to third-body perturbations. We also consider six other detector pointings (P1-6) spreading over the mean lunar orbital plane, and present the corresponding optimized orbits, which one may considered as backups or alternatives for the current design. Particularly, it allows a possibility of adding a second constellation of three spacecraft (e.g., P4) with the orbital plane roughly perpendicular to the first constellation. In this way, year-round GW observation can be arranged without interruption due to sunlight.

In future studies, employment of orbital corrections that can restrain relative motion between satellites for longer periods will be investigated. The maneuvers can take advantage of the semi-annual transition periods (3 months or less) when sunlight is roughly aligned with the orbital plane and interferes with GW observation \cite{Luo2016}. Furthermore, the requirement on delivery accuracy into the target orbits is being worked out, and the preliminary estimation has shown promise. Combined schemes, such as inter-satellite laser ranging, Chinese Deep Space Network, BeiDou/GPS, satellite laser ranging, etc., will be considered to assess the future orbit determination capability. 


\section*{Acknowledgements} 
The authors thank Gang Wang, Shoucun Hu, Yi-Ming Hu, and Hsien-Chi Yeh for helpful discussion. Our gratitude extends to the developers of GMAT. The work is supported by NSFC 11805287, 91636111, 11690022, 11475064, 11690021, 11503007, and 41274041.


\appendix

\section{Estimation on eccentricities} \label{app:ecc}

To provide some intuition, one can use a two-body model (see also \cite{Hu2018}) to roughly estimate the stability requirements on the eccentricities of the spacecraft orbits. To begin with, we assume that the Keplerian orbits in the same orbital plane are given by
\begin{eqnarray}
X_{k} = a(\cos\psi_{k}-e), \qquad 
Y_{k} = a\sqrt{1-e^{2}} \sin\psi_{k}, \qquad k=1,2,3. \label{1}
\end{eqnarray}
The eccentric anomaly $\psi_k$ is determined from the mean anomaly $\text{M}_k$ and the eccentricity $e$:
\begin{eqnarray}
\psi _{k}-e\sin \psi _{k}= \text{M}_{k},\qquad \text{M}_{k}\equiv n\,  t-\text{M}_{k0}, \label{2}
\end{eqnarray}
with the mean angular motion $n$, the time $t$, and a constant $\text{M}_{k0}$. Thereby the eccentric anomaly can be obtained as
\begin{eqnarray}
\psi_{k}=\text{M}_{k}+e \sin \text{M}_{k}+e^{2}\cos \text{M}_{k} \sin \text{M}_{k}+\mathcal{O}(e^{3}). \label{3}
\end{eqnarray}
Moreover, we have
\begin{eqnarray}
\text{M}_{k}=n\, t + \frac{2\pi}{3}(k-1), \label{4}
\end{eqnarray}
which follows from assuming equal arm lengths $ L_{12}=L_{13}=L_{23} $ when $e=0$. For these elliptic orbits, we can infer that the formation stability is closely related to the eccentricity by the following relations:
\begin{equation}
\eqalign{\left | \Delta l_{ij}(t) \right |_{\max}\equiv \left | \frac{L_{ij}(t)-L_{0}}{L_{0}} \right |_{\max}\approx  \frac{1}{2}e\approx \frac{e}{0.02} \times 1\, (\%),   \cr
\left | v_{ij}(t) \right |_{\max}\equiv \left |\dot L_{ij}(t) \right |_{\max}\approx \frac{\sqrt{3} a n}{2}e \approx \sqrt{\frac{10^5}{a}}\frac{e}{0.0058}\times 10\, (\text{m/s}), \cr
\left | \Delta \alpha_{k}(t) \right |_{\max}\equiv \left | \alpha _{k}(t)-\frac{\pi }{3} \right |_{\max}\approx \frac{\sqrt{3}}{2}e \approx \frac{e}{0.004}\times 0.2\, (^{\circ}), } \label{5} 
\end{equation}
with $ L_{0}=\sqrt{3}\,a $. The estimate $e\sim 10^{-3}$, as an upper bound, is consistent with our optimized orbits from previous sections. 


\section{Derivation of equations \eref{eq_rv}} \label{app:eq_rv}

To remove long-term drift in arm lengths and breathing angles, one needs to make sure that the mean angular velocities, or equivalently, the mean semi-major axes by Kepler's law, are the same for the three spacecraft (see, e.g., \cite{Men2010,Wan2017}).

For real-world, perturbed orbits at a given moment, the Keplerian description can still apply. A perturbed orbit can be described by a set of mean elements $ \bar{\sigma} $, together with secular terms $ \sigma_{c} $ and periodic terms $ \sigma_{ls} $ (including both long-periodic and short-periodic terms) \cite{Vallado1997,Liu2000}:
\begin{eqnarray}
\sigma (t)=\bar{\sigma }(t)+ \sigma _{ls}(t), \quad \bar{\sigma }(t)=\bar{\sigma }_{0}+ \eta\,  \bar{n}\,(t-t_{0})+\sigma _{c}(t), \quad \bar{\sigma }_{0}=\sigma _{0}- \sigma _{ls}(t_{0}),\label{9}
\end{eqnarray}
where $ \sigma \in \{ a, e, i,  \Omega, \omega, \text{M} \} $, and $ \eta=1 $ if $\sigma=\text{M}$, and $ \eta=0 $ if $\sigma\neq\text{M}$, and $ \bar{\sigma }_{0} $ denotes the initial mean elements.

On the form of $ \sigma (t) $, the secular terms are composed of linear functions or polynomials of $ (t-t_{0}) $, and the long-periodic terms of trigonometric functions of ($ \Omega $, $ \omega $), and the short-periodic terms of trigonometric functions of $ \text{M} $. In addition, their coefficients are functions of ($ a$, $e $, $i $)~\cite{Liu2000} (Liu 2000, pp 96, 126, 268). For example, we have
\begin{eqnarray}
a _{ls}(t)\sim F_{1}(a,e,i)\cdot F_{2}(\Omega ,\omega ,\text{M}), \label{10}\\
\Omega_{c} (t)\sim F_{3}(a,e,i)\cdot (t-t_{0}), \label{11} 
\end{eqnarray}
with some functions $ F_{1} $, $ F_{2} $, and $ F_{3} $ (also $ F_{4}(e,i,\Omega ,\omega ,\text{M}) $ in equation \eref{14}).

In the case of conservative perturbations, one has $ a_{c}(t)=0 $ for the semi-major axis $ a(t) $ \cite{Liu2000,Vallado1997,Smith1962} (Liu 2000, p 117; Vallado 1997, p 588). From equation \eref{9}, we obtain the mean element $ \bar{a}(t) $ as
\begin{eqnarray}
\bar{a }(t)=a _{0}- a _{ls}(t_{0}).\label{12}
\end{eqnarray}
Taking an variation of \eref{12}, we arrive at
\begin{eqnarray}
\delta \bar{a }=\delta a _{0}-\delta a _{ls}(t_{0}) \label{13}
\end{eqnarray}
with $ \delta a _{0}=a_{0}'- a _{0} $. From perturbative analysis \cite{Liu2000} (Liu2000, pp 126, 268), one has
\begin{eqnarray}
a _{ls}(t) \sim a_{0}^{\gamma }F_{4}(e,i,\Omega ,\omega ,\text{M}),\label{14}
\end{eqnarray}
where $ \gamma=-1 $ for the Earth's $ J_{2} $ perturbation and $ \gamma=4 $ for the lunisolar perturbations. Because the lunisolar effects are far greater in magnitude than the $ J_{2} $ perturbation, we adopt the relation $ a _{ls}(t) \approx  a_{0}^{4 }F_{4}(e,i,\Omega ,\omega ,\text{M}) $. Therefore, we have
\begin{eqnarray}
\delta a _{ls}(t_{0})=\frac{\partial a _{ls}}{\partial a_{0}} \delta a _{0} \approx  4 \frac{a _{ls}(t_{0})}{a _{0}}\delta a _{0},\label{15}
\end{eqnarray}
assuming that all other elements are fixed. One can regard $ \bar{a} $ as a perturbation of $ a_{0} $:
\begin{eqnarray}
\bar{a}=a_{0}+\varepsilon a_{0},\label{16}
\end{eqnarray}
with $ \left | \varepsilon  \right | \sim 10^{-4} $ for a geocentric spacecraft orbiting at an altitude of $ \sim 10^{5} $ km. Substituting equations \eref{12} and \eref{16} into equation \eref{15}, we obtain
\begin{eqnarray}
\delta a _{ls}(t_{0}) \approx  -4 \varepsilon \delta a _{0}, \label{17}
\end{eqnarray}
and by equation \eref{13},
\begin{eqnarray}
\delta \bar{a } \approx (1+4 \, \varepsilon)\delta a _{0}. \label{18}
\end{eqnarray}
Combining \eref{16} and \eref{18}, we find
\begin{eqnarray}
\frac{\delta a _{0}}{a _{0}} \approx \frac{1+\varepsilon }{1+4 \, \varepsilon }\frac{\delta \bar{a }}{\bar{a }},\qquad 
a_{0}' \approx \left ( 1+\frac{1+\varepsilon }{1+4 \, \varepsilon } \frac{{\bar{a }}'-\bar{a }}{\bar{a }} \right )a _{0},\label{19}
\end{eqnarray}
with $ \varepsilon = (\bar{a}-a_{0})/a_{0} $ from \eref{16}. 

In practice, we would prefer using the equation \eref{19} in terms of the spacecraft's position and velocity, as in \cite{Men2010,Wang2013b}. As mentioned before, geometric relations in elliptical motion can still apply to a perturbed Keplerian orbit, e.g., 
\begin{eqnarray}
r=a (1-e \cos \psi ),\label{20}
\end{eqnarray}
where each quantity now varies with time. Considering the case with $ \delta e = 0$ and $ \delta \text{M} = 0$, we obtain from \eref{20} 
\begin{eqnarray}
\frac{\delta  r}{r}=\frac{\delta a }{a}. \label{21}
\end{eqnarray}
For perturbed Keplerian orbits, the kinetic and potential energy is related to the semi-major axis by
\begin{eqnarray}
\frac{m v^2}{2}-\frac{G M_{\oplus } m  }{r}=-\frac{G M_{\oplus } m }{2 a} \label{22}
\end{eqnarray}
at each given time. Taking the variation of \eref{22} gives
\begin{eqnarray}
v \delta v=\frac{G M_{\oplus } }{2 a^2}\delta a - \frac{G M_{\oplus } }{r^2}\delta r.\label{23}
\end{eqnarray}
Combining equations \eref{21}, \eref{22} and \eref{23}, we obtain
\begin{eqnarray}
\frac{\delta v}{v}=-\frac{\delta a }{2a}.\label{24}
\end{eqnarray}
By equations \eref{19}, \eref{21} and \eref{24}, we find
\begin{eqnarray}
\frac{\delta \bi{r}_{0}}{\bi{r}_{0}}=\frac{\delta  r_{0}}{r_{0}}=\frac{\delta a_{0} }{a_{0}}\approx \frac{1+\varepsilon }{1+4 \, \varepsilon } \frac{\delta \bar{a }}{\bar{a }},\qquad \frac{\delta \bi{v}_{0}}{\bi{v}_{0}}\approx - \frac{1+\varepsilon }{1+4 \, \varepsilon } \frac{\delta \bar{a }}{2\bar{a }}, \label{25}
\end{eqnarray}
or,
\begin{eqnarray}
\bi{r}'_{0}\approx \left ( 1+\frac{1+\varepsilon }{1+4 \, \varepsilon } \frac{{\bar{a }}'-\bar{a }}{\bar{a }} \right )\bi{r}_{0},\qquad \bi{v}'_{0}\approx \left ( 1- \frac{1+\varepsilon }{1+4 \, \varepsilon } \frac{{\bar{a }}'-\bar{a }}{2\bar{a }} \right )\bi{v}_{0}, \label{26}
\end{eqnarray}
which completes the derivation of equations \eref{eq_rv}.


\section{Derivation of equations \eref{eq_io}} \label{app:eq_io}

We also adjust the orbital planes of the three spacecraft so that they can stay in the same average plane. The inclination $ i $ and the longitude of ascending node $ \Omega $ determine the orientation of the orbital plane. For the inclination, it has little secular change. Hence, similar to the treatment of $a$, we apply the same form of the iteration \eref{19} on $ i $, i.e.,
\begin{eqnarray}
i _{0}' \approx \left ( 1+ \frac{1+\epsilon  }{1+4 \, \epsilon } \frac{\bar{i }^{\, '}-\bar{i }}{\bar{i }} \right )i _{0},\label{27}
\end{eqnarray}
with the average inclination $ \bar{i} $, the desired inclination $ \bar{i }^{\, '} $, and $ \epsilon = (\bar{i}-i_{0})/i_{0} $. For the longitude of ascending node, we know from \eref{9} that
\begin{eqnarray}
\bar{\Omega}(t) =\Omega_{0}-\Omega_{ls} (t_{0})+\Omega_{c} (t).\label{28}
\end{eqnarray}
Taking an average over time $ T $, one has
\begin{eqnarray}
\bar{\Omega}_{T}\equiv \frac{1}{T}\int_{t_{0}}^{t_{0}+T}\bar{\Omega}(t) \, dt \approx \Omega_{0}+\frac{1}{T}\int_{t_{0}}^{t_{0}+T}\Omega_{c} (t)\, dt,\label{29}
\end{eqnarray}
where the small contribution from $\Omega_{ls} (t_{0})$ is neglected. By the relation \eref{11} it follows that $ \delta \bar{\Omega}_{T}\approx \delta \Omega_{0} $. Using $ \bar{\Omega}(t)=\Omega(t)-\Omega_{ls}(t) $ from \eref{9}, we arrive at $ \bar{\Omega} \equiv \frac{1}{T}\int_{t_{0}}^{t_{0}+T} \Omega(t) \, dt = \bar{\Omega}_{T} $ and
\begin{eqnarray}
\Omega_{0}^{'} \approx \Omega_{0}+\left (\bar{\Omega}^{'} -\bar{\Omega} \right ),\label{30}
\end{eqnarray}
where $ \bar{\Omega}^{'} $ is the desired mean longitude of ascending node.


\section*{References}

\begin{thebibliography}{99}

\bibitem{Luo2016}
Luo J {\it et al} 2016 TianQin: a space-borne gravitational wave detector {\it Class. Quantum Grav.} {\bf 33} 035010

\bibitem{Stroeer2006} 
Stroeer A and Vecchio A 2006 The LISA verification binaries, {\it Class. Quantum Grav.} {\bf 23} S809--17

\bibitem{Hu2017}
Hu Y-M, Mei J-W and Luo J 2017 Science prospects for space-borne gravitational-wave missions {\it National Science Review} {\bf 4} 683--4

\bibitem{Montenbruck2001}  
Montenbruck O and Gill E 2001 {\it Satellite Orbits: Models, Methods, and Applications} (New York: Springer) p~114

\bibitem{Vincent1987}
Vincent M A and Bender P L 1987 {\it Proc. Astrodynamics Specialist Conf. (Kalispell)} vol 1 (San Diego, CA: Univelt) p 1346

\bibitem{Folkner1997}
Folkner W M, Hechler F, Sweetser T H, Vincent M A and Bender P L 1997 LISA orbit selection and stability {\it Class. Quantum Grav.} {\bf 14} 1405--10

\bibitem{Sweetser2005}
Sweetser T H 2005 An end-to-end trajectory description of the LISA mission {\it Class. Quantum Grav.} {\bf 22} S429--35

\bibitem{Dhurandhar2005}
Dhurandhar S V, Nayak K R, Koshti S and Vinet J-Y 2005 Fundamentals of the LISA stable flight formation {\it Class. Quantum Grav.} {\bf 22} 481--7

\bibitem{Nayak2006}
Nayak K R, Koshti S, Dhurandhar S V and Vinet J-Y 2006 On the minimum flexing of LISA's arms {\it Class. Quantum Grav.} {\bf 23} 1763--78

\bibitem{Marchi2012}
De Marchi F, Pucacco G and Bassan M  2012 Optimizing the Earth–-LISA `rendezvous' {\it Class. Quantum Grav.} {\bf 29} 035009

\bibitem{Yi2008}
Yi Z-H, Li G-Y, Heinzel G, R\"{u}diger A, Jennrich O, Wang L, Xia Y, Zeng F and Zhao H-B 2008 Coorbital restricted problem and its application in the design of the orbits of the LISA spacecraft {\it Int. J. Mod. Phys.} D {\bf 17} 1005--19

\bibitem{Hughes2002}
Hughes S P 2002 Preliminary optimal orbit design for laser interferometer space antenna {\it 25th Annual AAS Guidance and Control Conference (Breckenridge CO, Feb. 2002)}

\bibitem{Povoleri2006} 
Povoleri A and Kemble S 2006 LISA orbits {\it AIP Conf. Proc.} {\bf 873} 702–-6

\bibitem{Li2008}
Li G-Y, Yi Z-H, Heinzel G, R\"{u}diger A, Jennrich O, Wang L, Xia Y, Zeng F and Zhao H-B 2008 Methods for orbit optimization for the LISA gravitational wave observatory {\it Int. J. Mod. Phys.} D {\bf 17} 1021--42

\bibitem{Xia2010}
Xia Y, Li G-Y, Heinzel G, R\"udiger A and Luo Y-J 2010 Orbit design for the laser interferometer space antenna (LISA) {\it Sci. China Phys. Mech. Astron.} {\bf 53} 179--86

\bibitem{Halloin2017}
Halloin H 2017 Optimizing orbits for (e)LISA {\it J. Phys.: Conf. Ser.} {\bf 840} 012048

\bibitem{Wang2013a}
Wang G and Ni W-T 2013 Numerical simulation of time delay interferometry for eLISA/NGO {\it Class. Quantum Grav.} {\bf 30} 065011

\bibitem{Ni2013}
Ni W-T 2013 ASTROD-GW: overview and progress {\it Int. J. Mod. Phys.} D {\bf 22} 1341004

\bibitem{Men2010}
Men J-R, Ni W-T and Wang G 2010 Design of ASTROD-GW Orbit {\it Chin. Astron. Astrophys.} {\bf 34} 434--46

\bibitem{Wang2013b}
Wang G and Ni W-T 2013 Orbit optimization for ASTROD-GW and its time delay interferometry with two arms using CGC ephemeris {\it Chin. Phys.} B {\bf 22} 049501

\bibitem{Wang2015}
Wang G and Ni W-T 2015 Orbit optimization and time delay interferometry for inclined ASTROD-GW formation with half-year precession-period {\it Chin. Phys.} B {\bf 24} 059501

\bibitem{Ni2016}
Ni W-T 2016 Gravitational wave detection in space {\it Int. J. Mod. Phys.} D {\bf 25} 1630001

\bibitem{Hu2015} 
Hu S-C, Zhao Y-H and Ji J-H 2015 Internal report by Purple Mountain Observatory (unpublished)

\bibitem{GMAT}
\href{http://gmatcentral.org/}{http://gmatcentral.org}

\bibitem{Hughes14}
Hughes S P, Qureshi R H, Cooley D S, and Parker J J 2014 Verification and Validation of the General Mission Analysis Tool (GMAT), {\it AIAA/AAS Astrodynamics Specialist Conference, AIAA SPACE Forum}, (AIAA 2014-4151) 

\bibitem{Tapley1996}	
Tapley B D {\it et al} 1996 The Joint Gravity Model 3 {\it J. Geophys. Res.} {\bf 101} 28029--49

\bibitem{Folkner2008}
Folkner W M, Williams J G and Boggs D H 2008 The Planetary and Lunar Ephemeris DE 421, Memorandum IOM 343R-08-003, Jet Propulsion Laboratory, California Institute of Technology

\bibitem{Wan2017}
Wan X-B, Zhang X-M and Li M 2017 Analysis of long-period drift characteristics for orbit configuration of the TianQin Mission (in Chinese) {\it Chinese Space Science and Technology} {\bf 37} 110--6

\bibitem{Smith1962}
Smith D E 1962 The perturbation of satellite orbits by extra-terrestrial gravitation {\it Planet. Space Sci.} {\bf 9} 659--74

\bibitem{Vallado1997}
Vallado D A 1997 {\it Fundamentals of Astrodynamics and Applications} (New York: McGraw Hill)

\bibitem{Liu2000}
Liu L 2000 {\it Orbit Theory of Spacecraft} (in Chinese) (Beijing: National Defence Industry Press)

\bibitem{JPL}
\href{https://ssd.jpl.nasa.gov/horizons.cgi}{https://ssd.jpl.nasa.gov/horizons.cgi}

\bibitem{Hu2018}
Hu X-C, Li X-H, Wang Y, Feng W-F, Zhou M-Y, Hu Y-M, Hu S-C, Mei J-W and Shao C-G 2018 Fundamentals of the orbit and response for TianQin {\it Class. Quantum Grav.} {\bf 35} 095008

\end{thebibliography}

\end{document}